\documentclass[aps,prl,reprint,superscriptaddress,longbibliography]{revtex4-2}
\usepackage[T1]{fontenc}
\usepackage{amsmath,amssymb,amsthm}
\usepackage{mathtools,booktabs,longtable,tikz,microtype}
\usepackage{xcolor}
\definecolor{apsblue}{rgb}{0.18, 0.19, 0.57}
\usepackage[colorlinks=true,allcolors=apsblue, breaklinks=true]{hyperref}

\usetikzlibrary{arrows.meta}

\makeatletter
\DeclareFontFamily{OMX}{MnSymbolE}{}
\DeclareSymbolFont{MnLargeSymbols}{OMX}{MnSymbolE}{m}{n}
\SetSymbolFont{MnLargeSymbols}{bold}{OMX}{MnSymbolE}{b}{n}
\DeclareFontShape{OMX}{MnSymbolE}{m}{n}{
    <-6>  MnSymbolE5
   <6-7>  MnSymbolE6
   <7-8>  MnSymbolE7
   <8-9>  MnSymbolE8
   <9-10> MnSymbolE9
  <10-12> MnSymbolE10
  <12->   MnSymbolE12
}{}
\DeclareFontShape{OMX}{MnSymbolE}{b}{n}{
    <-6>  MnSymbolE-Bold5
   <6-7>  MnSymbolE-Bold6
   <7-8>  MnSymbolE-Bold7
   <8-9>  MnSymbolE-Bold8
   <9-10> MnSymbolE-Bold9
  <10-12> MnSymbolE-Bold10
  <12->   MnSymbolE-Bold12
}{}

\let\llangle\@undefined
\let\rrangle\@undefined
\DeclareMathDelimiter{\llangle}{\mathopen}%
                     {MnLargeSymbols}{'164}{MnLargeSymbols}{'164}
\DeclareMathDelimiter{\rrangle}{\mathclose}%
                     {MnLargeSymbols}{'171}{MnLargeSymbols}{'171}
\makeatother

\begin{document}
\title{Hidden Geometric Structure Beyond Conformal Flatness}
\author{Xinan Zhou}
\affiliation{Kavli Institute for Theoretical Sciences, University of
Chinese Academy of Sciences, Beijing 100190, China}
\date[]{} 
\begin{abstract}
We uncover a hidden geometric structure in holographic correlators of M-theory, despite the absence of conformal flatness. We study two-point functions of $\frac{1}{2}$-BPS local operators in the presence of a $\frac{1}{2}$-BPS surface defect in six-dimensional $(2,0)$ theory, dual to an M2-brane in $\mathrm{AdS}_7\times$S$^4$. We find that the infinite tower of tree-level supergravity correlators can be generated from a single lowest-weight correlator. After a simple reorganization of the rational terms, the generating function takes the form of a weighted average over the internal S$^4$, acting through additive shifts of the spacetime cross ratios. This reveals a remarkably simple geometric organization underlying the holographic defect correlators.
\end{abstract}
\maketitle

\section{Introduction}
Over the past decade, bootstrap techniques \cite{Rastelli:2016nze,Rastelli:2017udc} have allowed holographic correlators to be efficiently computed in a wide variety of theories, producing systematic results that go far beyond the traditional diagrammatic algorithm (see \cite{Zhou:2026xtp} for a recent review). Besides their simplicity, the results exhibit various interesting hidden structures. One particularly remarkable structure is the higher dimensional hidden conformal symmetry. First observed in supergravity four-point functions in AdS$_5\times$S$^5$ \cite{Caron-Huot:2018kta}, it packages infinitely many correlators into a single generating function. The same structure was later found in AdS$_3\times$S$^3$ supergravity \cite{Rastelli:2019gtj,Giusto:2020neo}, and even in the non-gravitational theory of $\mathcal{N}=1$ SYM in AdS$_5\times$S$^3$ \cite{Alday:2021odx}. Hidden conformal symmetry has also proved particularly useful in simplifying the computation of holographic correlators at higher loops and higher points \cite{Caron-Huot:2018kta,Aprile:2019rep,Alday:2019nin,Huang:2023ppy,Huang:2024dxr,Fernandes:2025eqe}, and in revealing an AdS double copy structure \cite{Zhou:2021gnu}.

More recently, a partially broken version of hidden conformal symmetry was found in four-point functions of two light supergravitons and two heavy maximal giant gravitons \cite{Chen:2025yxg,Chen:2026ium},  where the heavy operator pair introduces a defect breaking the higher dimensional symmetry. This structure has been further extended to non-maximal and dual giant gravitons \cite{Chen:2026yzm}. The simplification there is especially striking. Direct Witten diagram calculations and bootstrap lead to complicated expressions, whereas hidden conformal symmetry allows the generating functions to be obtained from the known maximal giant-graviton result by simple $SO(2,10)$ rotations.

It is a good moment to take stock and think about what all these examples have in common. A shared feature is that infinitely many correlators are generated from a single lowest-weight reduced correlator. Moreover, the generating functions can be obtained by simple uplifting: the spacetime distances are promoted to higher dimensional conformal invariants incorporating the R-symmetry polarizations. In the simplest cases, this amounts to replacing $x_{ij}^2$ by $x_{ij}^2-2t_i\cdot t_j$, which naturally arises as an inner product of higher dimensional null vectors. The geometric origin of this prescription is closely related to the conformal flatness of the relevant $\mathrm{AdS}\times$S backgrounds. 

The success of such a construction makes it tempting to regard conformal flatness as an essential ingredient of the hidden simplicity. We find, however, that a single-seed structure survives beyond conformal flatness, in a qualitatively new form. We consider six-dimensional $(2,0)$ theory, dual to eleven-dimensional supergravity on AdS$_7\times$S$^4$, where the unequal radii of the two factors prevent conformal flatness. Our observables are two-point functions of $\frac{1}{2}$-BPS local operators in the presence of a $\frac{1}{2}$-BPS surface defect, dual to an M2-brane occupying AdS$_3$. Although these correlators have been computed for arbitrary external weights using bootstrap techniques \cite{Chen:2023yvw}, the known expressions reveal no obvious hidden structure. Remarkably, after a simple reorganization of the rational terms, the entire tower of tree-level reduced correlators can again be generated from a single lowest-weight correlator. The generating function, however, is no longer a direct higher-dimensional uplift of spacetime distances. Instead, it involves a weighted average over the internal S$^4$, with the seed evaluated at shifted distances. Explicitly, defining $\mathbf{H}=\sum_{k_1,k_2\geq 2}s^{k_1}t^{k_2}h_{k_1k_2}$, our central result reads
\begin{equation}\label{genHintro}
\mathbf{H}=\frac{4s^2t^2}{3L^3}\!
\int_{{\rm S}^4}\!\frac{d\Omega_4(\vartheta)}{\operatorname{Vol}({\rm S}^4)}|\Pi_{E^\perp}\vartheta|\frac{D_-}{\widetilde D_-}h_{22}\left(\frac{\widetilde D_-}{\sqrt L},\frac{\widetilde D_+-\widetilde D_-}{2\sqrt L}\right).
\end{equation}
Here $L=(1-s)(1-t)$ and $\widetilde{D}_\pm$ are shifted cross ratios which depend on $s$, $t$ and the internal integration point $\vartheta$. Their precise definitions will be given below.

\section{Defect two-point functions}
We consider $\frac{1}{2}$-BPS local operators in 6d $(2,0)$ theory
\begin{equation}
\mathcal{O}_k(x,u)=\mathcal{O}^{I_1\cdot I_k}u_{I_1}\ldots u_{I_k}\;, \quad u\cdot u=0\;,
\end{equation}
which have protected conformal dimension $2k$. The $SO(5)$ null vector $u$ projects the operator to the rank-$k$ symmetric traceless representation of R-symmetry. Adding a $\frac{1}{2}$-BPS surface defect selects a subspace $\mathbb{R}^2\subset \mathbb{R}^6$. This separates $x$ into a parallel part and a transverse part
\begin{equation}
x_i=(y_i,z_i)\;,\qquad y_i\in\mathbb{R}^2\;,\qquad z_i\in\mathbb{R}^4\;.
\end{equation}
The defect breaks the conformal group $SO(1,7)$ to $SO(1,3)\times SO(4)$, and the $SO(5)$ R-symmetry to $SO(4)$. The latter can be most conveniently seen from the holographic realization of the surface defect. The $\frac{1}{2}$-BPS defect corresponds to a single M2 brane which is embedded as an AdS$_3$ subspace inside of AdS$_7$ sitting at a point on the internal S$^4$. The S$^4$ location is described by a fixed unit vector $\theta\in{\rm S}^4$ with $\theta^2=1$, which leaves only $SO(4)\subset SO(5)$ invariant. We focus on the two-point functions of $\frac{1}{2}$-BPS operators in the presence of the surface defect. The defect correlator can be written as
\begin{equation}
\llangle \mathcal{O}_{k_1}\mathcal{O}_{k_2}\rrangle=\frac{(u_1\cdot \theta)^{k_1}(u_2\cdot \theta)^{k_2}}{(z_1^2)^{k_1}(z_2^2)^{k_2}}\mathcal{F}(z,\bar{z},\sigma)\;,
\end{equation}
where the conformal cross ratios are 
\begin{equation}
\nonumber\frac{x_{12}^2}{|z_1||z_2|}=\frac{(1-z)(1-\bar{z})}{\sqrt{z\bar{z}}}=\xi\;,\quad \frac{z_1\cdot z_2}{|z_1||z_2|}=\frac{z+\bar{z}}{2\sqrt{z\bar{z}}}=\frac{\chi}{2}\;,
\end{equation}
and the R-symmetry cross ratio is
\begin{equation}
\nonumber \sigma=\frac{u_1\cdot u_2}{(u_1\cdot \theta)(u_2\cdot \theta)}=-\frac{(1-\omega)^2}{2\omega}\;.
\end{equation}
The preserved supercharges further impose superconformal Ward identities which relate the $z$, $\bar{z}$ and $\omega$ dependence \cite{Meneghelli:2022gps}
\begin{equation}
\partial_{\bar{z}}\left(\mathcal{F}_{k_1k_2}\big|_{\omega=\bar{z}}\right)=0\;,\quad \partial_z\left(\mathcal{F}_{k_1k_2}\big|_{\omega=z}\right)=0\;.
\end{equation}
Their solution reads 
\begin{equation}
\mathcal{F}_{k_1k_2}=\mathcal{F}_{{\rm rat},k_1k_2}+{\rm R}\, \mathcal{H}_{k_1k_2}\;,
\end{equation}
where ${\rm R}$ is determined by superconformal symmetry
\begin{equation}
{\rm R}=(z\bar{z})^{-1}(z-\omega)(\bar{z}-\omega)(z-\omega^{-1})(\bar{z}-\omega^{-1})\;.
\end{equation}
The correlator is therefore divided into a rational part  $\mathcal{F}_{{\rm rat},k_1k_2}$ with the property
\begin{equation}
\mathcal{F}_{k_1k_2}\big|_{\omega=\bar{z}}=g_{k_1k_2}(z)\;,\quad \mathcal{F}_{k_1k_2}\big|_{\omega=z}=g_{k_1k_2}(\bar{z})\;,
\end{equation}
and a reduced correlator $\mathcal{H}_{k_1k_2}$ containing all the dynamical information. 

These defect correlators were systematically computed in \cite{Chen:2023yvw} for arbitrary $k_1$, $k_2$ using a bootstrap strategy. Schematically, \cite{Chen:2023yvw} makes an ansatz for the correlator as a sum of possible Witten diagrams with unfixed coefficients and then uses superconformal Ward identities to determine these coefficients. The general reduced correlators were given in the defect Mellin representation \cite{Rastelli:2017ecj,Goncalves:2018fwx}
\begin{equation}\label{Mellinrep}
\begin{split}
\mathcal{H}_{k_1k_2}={}&\int \frac{d\delta d\gamma}{(2\pi i)^2}\xi^{-\delta}\chi^{-\gamma+\delta}\widetilde{\mathcal{M}}_{k_1k_2}(\delta,\gamma,\sigma)\\
 {}& \times\Gamma(\delta+1)\Gamma(\gamma-\delta)\prod_{i=1}^2\Gamma(k_i+1-\frac{\gamma}{2})\;,
\end{split}
\end{equation}
where we close the $\delta$ contour to the left to circle the exchange poles and the $\gamma$ contour to the right.
Here it is useful to rewrite the Mellin amplitudes of \cite{Chen:2023yvw} as
\begin{equation}\label{Mtildek1k2}
\frac{\widetilde{\mathcal{M}}_{k_1k_2}}{b_{k_1\mathcal{D}}b_{k_2\mathcal{D}}}=\sum_{j=2}^{k_{\rm m}}\frac{K_j}{(k_1-j)!(k_2-j)!}\;,
\end{equation}
where $k_{\rm m}=\min\{k_1,k_2\}$ and $b_{k\mathcal{D}}$ are bulk-defect OPE coefficients. The numerator $K_j$ has no dependence on $k_1$ and $k_2$, and reads 
\begin{equation}\label{defKj}
K_j=\frac{1}{\gamma-2j}\sum_{i=1}^{2j-2}\frac{r_{ij}(\sigma)}{\delta-i}\;,
\end{equation}
with 
\begin{equation}
r_{ij}(\sigma)=\sum_{m=\lfloor \frac{i}{2}\rfloor}^{\min\{i,j-1\}}\frac{(-1)^i\binom{m}{i-m}(2\sigma)^{m-1}}{2j!m!(j-m-1)!}\;.
\end{equation}
For later convenience, we define
\begin{equation}
h_{k_1k_2}=\frac{\mathcal{H}_{k_1k_2}}{b_{k_1\mathcal{D}}b_{k_2\mathcal{D}}}\;.
\end{equation}
The meromorphic functions were also obtained in \cite{Chen:2023yvw}
\begin{equation}\label{meromorphg}
g_{k_1k_2}(z)=\frac{1}{2}b_{k_1\mathcal{D}}b_{k_2\mathcal{D}}\sum_{i=1}^{k_{\rm m}-1}C_i\left(-\frac{z}{(1-z)^2}\right)^i\;,
\end{equation}
where $C_i=\frac{1}{i+1}\binom{2i}{i}$ is the Catalan number. In \cite{Chen:2023yvw}, the rational part was defined in terms of the meromorphic function as  
\begin{equation}
\mathcal{F}_{{\rm rat},k_1k_2}^{{\rm old}}=\frac{(z-\omega)(z-\omega^{-1})}{(z-\bar{z})(z-\bar{z}^{-1})}g_{k_1k_2}(z)+(z\leftrightarrow\bar{z})\;.
\end{equation}
In this paper, we will use instead a different definition 
\begin{equation}
\mathcal{F}_{{\rm rat},k_1k_2}=\frac{b_{k_1\mathcal{D}}b_{k_2\mathcal{D}}}{2}\sum_{i=1}^{k_{\rm m}-1}C_i \left(\frac{2\sigma}{\xi^2}\right)^i\;.
\end{equation}
The difference takes the form
\begin{equation}
\mathcal{F}_{{\rm rat},k_1k_2}=\mathcal{F}_{{\rm rat},k_1k_2}^{{\rm old}}+b_{k_1\mathcal{D}}b_{k_2\mathcal{D}}{\rm R}\, Q_{k_{\rm m}}\;,
\end{equation}
where $Q_{k_{\rm m}}$ is rational and depends on the external weights only through $k_{\rm m}$. It is nonvanishing only for $k_{\rm m}>2$. Note this amounts to a reshuffling between $\mathcal{F}_{{\rm rat},k_1k_2}$ and $\mathcal{H}_{k_1k_2}$, i.e., $h_{k_1k_2}=h_{k_1k_2}^{\rm old}-Q_{k_{\rm m}}$. The rationality of $Q_{k_{\rm m}}$ makes the redefinition invisible in terms of the Mellin amplitude \cite{Rastelli:2017udc}. However, the difference is important in position space. The protected part  $\mathcal{F}_{{\rm rat},k_1k_2}$ is significantly simpler compared to \cite{Chen:2023yvw}, and upon setting $\omega=\bar{z}$ manifestly leads to (\ref{meromorphg}). More importantly, the reduced correlators $\mathcal{H}_{k_1k_2}$ defined in this way turn out to admit a hidden geometric structure which allows them to be organized into the generating function (\ref{genHintro}).  For the lowest charges, $Q_2=0$, and the normalized seed takes the explicit form
\begin{equation}
h_{22}=-\frac{1}{20\xi^2}\,{}_2F_1\!\left(2,4;\frac{7}{2};r\right)-\frac{1}{70\xi}\,{}_2F_1\!\left(3,5;\frac{9}{2};r\right)\;,
\end{equation}
with $r=(2-\xi-\chi)/4$.

\section{Geometry intermezzo}
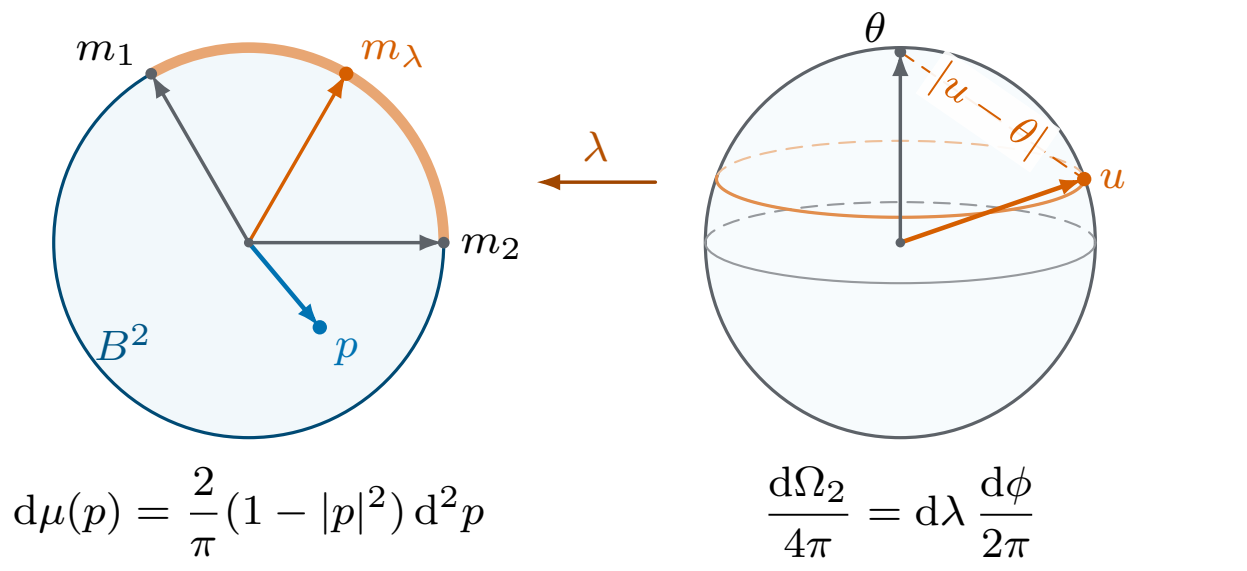
\begin{figure}[t]
\centering
\begin{tikzpicture}[
  x=1cm,y=1cm,
  font=\fontsize{8}{9.5}\selectfont,
  line cap=round,line join=round,
  >={Latex[length=1.6mm,width=1.1mm]},
  vector/.style={->,line width=.65pt},
  boundary/.style={line width=.6pt},
  guide/.style={line width=.4pt,densely dashed},
  every node/.style={inner sep=1pt}
]
\definecolor{geomblue}{HTML}{0072B2}
\definecolor{geomorange}{HTML}{D55E00}
\definecolor{geomgray}{HTML}{5D6268}

\path[use as bounding box] (0,-2.40) rectangle (8.40,1.62);

\begin{scope}[shift={(2.04,0)}]
  \def\rad{1.29}
  \coordinate (O) at (0,0);
  \coordinate (Mtwo) at (\rad,0);
  \coordinate (Mone) at (120:\rad);
  \coordinate (Mlambda) at (60:\rad);
  \coordinate (P) at (.47,-.56);

  \fill[geomblue!5] (O) circle (\rad);
  \draw[boundary,geomblue!65!black] (O) circle (\rad);
  \node[text=geomblue!75!black] at (-.83,-.66) {$B^2$};
  \draw[geomorange!55,line width=2pt]
    (0:\rad) arc[start angle=0,end angle=120,radius=\rad];
  \draw[vector,geomgray] (O)--(Mtwo);
  \draw[vector,geomgray] (O)--(Mone);
  \draw[vector,geomorange] (O)--(Mlambda);
  \draw[vector,geomblue,line width=.85pt] (O)--(P);
  \fill[geomgray] (O) circle (.9pt);
  \fill[geomgray] (Mtwo) circle (1.1pt);
  \fill[geomgray] (Mone) circle (1.1pt);
  \fill[geomorange] (Mlambda) circle (1.35pt);
  \fill[geomblue] (P) circle (1.35pt);
  \node[anchor=west] at (1.36,-.03) {$m_2$};
  \node[anchor=south east] at (-.69,1.13) {$m_1$};
  \node[anchor=south west,text=geomorange] at (.70,1.12) {$m_\lambda$};
  \node[anchor=north west,text=geomblue] at (.53,-.59) {$p$};
\end{scope}

\begin{scope}[shift={(6.35,0)}]
  \def\rad{1.29}
  \pgfmathsetmacro{\ce}{cos(12)}
  \pgfmathsetmacro{\se}{sin(12)}
  \pgfmathsetmacro{\zlat}{1/3}
  \pgfmathsetmacro{\rlat}{sqrt(1-\zlat*\zlat)}
  \pgfmathsetmacro{\latcut}{asin(\se*\zlat/(\ce*\rlat))}
  \coordinate (O) at (0,0);
  \coordinate (Theta) at (0,{\rad*\ce});
  \coordinate (U) at ({\rad*\rlat},{\rad*\ce*\zlat});

  \fill[geomblue!3] (O) circle (\rad);
  \draw[boundary,geomgray] (O) circle (\rad);
  \draw[guide,geomgray!65]
    (-\rad,0) arc[start angle=180,end angle=0,
      x radius=\rad,y radius={\rad*\se}];
  \draw[line width=.4pt,geomgray!65]
    (-\rad,0) arc[start angle=180,end angle=360,
      x radius=\rad,y radius={\rad*\se}];
  \draw[guide,geomorange!40]
    (0,{\rad*\ce*\zlat}) ellipse[
      x radius={\rad*\rlat},y radius={\rad*\se*\rlat}];
  \draw[line width=.6pt,geomorange!70]
    plot[domain={-\latcut}:{180+\latcut},samples=80,variable=\a]
    ({\rad*\rlat*cos(\a)},
     {\rad*(\ce*\zlat-\se*\rlat*sin(\a))});
  \draw[vector,geomgray] (O)--(Theta);
  \draw[vector,geomorange,line width=.85pt] (O)--(U);
  \draw[guide,geomorange] (Theta)--(U);
  \fill[geomgray] (O) circle (.9pt);
  \fill[geomgray] (Theta) circle (1.1pt);
  \fill[geomorange] (U) circle (1.35pt);
  \node[anchor=south east] at (-.05,{\rad*\ce+.02}) {$\theta$};
  \node[anchor=west,text=geomorange]
    at ({\rad*\rlat+.06},{\rad*\ce*\zlat}) {$u$};
  \node[fill=white,fill opacity=.92,text opacity=1,
        text=geomorange,rotate=-34.7]
    at (.58,.87) {$|u-\theta|$};
\end{scope}

\draw[->,geomorange!75!black,line width=.65pt]
  (4.73,.4)--(3.94,.4);
\node[text=geomorange!85!black,above] at (4.335,.5) {$\lambda$};
\node at (2.04,-1.78) {$\displaystyle \mathrm d\mu(p)=\frac{2}{\pi}(1-|p|^2)\,\mathrm d^2p$};
\node at (6.35,-1.78)   {$\displaystyle \frac{\mathrm d\Omega_2}{4\pi}
                      =\mathrm d\lambda\,\frac{\mathrm d\phi}{2\pi}$};
\node at (2.04,-2.45) {\textbf{(a)} Polarization plane $E$};
\node at (6.35,-2.45) {\textbf{(b)} Complementary $S^2\subset E^\perp$};

\end{tikzpicture}
\caption{Geometry of the internal average. (a) Weighted projection disk in $E$. (b) Unit sphere of complementary directions. The latitude of $u$ determines $\lambda$, and hence $m_\lambda$.}
\label{fig:geometry}
\end{figure}

To reveal the hidden structure in correlators, let us discuss the cross ratio geometry in more detail. We define
\begin{equation}
\frac{u_i}{(u_i\cdot \theta)}=\theta+{\rm i} m_i\;,\quad m_i\cdot \theta=0\;,\quad m_i^2=1\;,
\end{equation}
where $m_i$ are generally complex. But here we can restrict them to be real without loss of generality because the correlators depend on the R-symmetry polarizations only via the cross ratio $\sigma$. In terms of real $m_1$ and $m_2$, which are now in the tangent space $T_\theta {\rm S}^4$, we have
\begin{equation}
\sigma=1-m_1\cdot m_2=\frac{1}{2}(m_1-m_2)^2\in [0,2]\;.
\end{equation}
Since $\mathcal{F}_{k_1k_2}$ is analytic in $\sigma$, one can analytically continue in $\sigma$ when needed. The internal embedding space can be decomposed as $\mathbb{R}^5=E\oplus E^\perp$ where
\begin{equation}
E={\rm span}(m_1,m_2)\;,
\end{equation}
and $\theta\in E^\perp$. For a generic point $\vartheta\in S^4$, we define its projection to $E$ as
\begin{equation}\label{defp}
p=\Pi_E \vartheta\in E\;,\qquad |p|^2\leq 1\;.
\end{equation}
The projection to the transverse space is $\Pi_{E^\perp}\vartheta$ and has length $\sqrt{1-|p|^2}$. This gives
\begin{equation}
\vartheta=p+\sqrt{1-|p|^2}\, u\;,
\end{equation}
with $|u|=1$ and $u\in S^2\subset E^\perp$. Furthermore, we define
\begin{equation}\label{deflambda}
\lambda=\frac{1-\theta\cdot u}{2}=\frac{|u-\theta|^2}{4}\;, \quad 0\leq \lambda\leq 1\;.
\end{equation}
This allows us to define a unit vector $m_\lambda$ in the plane $E$ that interpolates continuously from $m_2$ to $m_1$, with its squared distance from $m_2$ increasing linearly
\begin{equation}\label{mlambdam2prod}
(m_\lambda-m_2)^2=\lambda(m_1-m_2)^2
=2\lambda\sigma\;.
\end{equation}
Equivalently, $m_\lambda\cdot m_2=1-\lambda\sigma$.  The geometry is illustrated in Figure \ref{fig:geometry}. Finally, we introduce for later convenience
\begin{equation}
D_\pm=\frac{y_{12}^2+(z_1\pm z_2)^2}{\sqrt{z_1^2z_2^2}}\;,
\end{equation}
in terms of which we can express $\xi$, $\chi$ as
\begin{equation}\label{xichitoDpm}
\xi=D_-\;,\quad \chi=\frac{D_+-D_-}{2}\;.
\end{equation}

\section{Level function and generator}
We now work from the other end by massaging the bootstrap result (\ref{Mtildek1k2}). We start by defining a simpler level function
\begin{equation}\label{defHj}
\begin{split}
H_j(\xi,\chi,\sigma)={}&\int \frac{d\delta d\gamma}{(2\pi i)^2}\xi^{-\delta}\chi^{-\gamma+\delta}K_j(\delta,\gamma,\sigma)\\
{}&\times\Gamma(\delta+1)\Gamma(\gamma-\delta)\Gamma^2(j+1-\frac{\gamma}{2})\;,
\end{split}
\end{equation}
and considering the sum 
\begin{equation}
\nonumber P_j(x,\sigma)=\sum_{i=1}^{2j-2}r_{ij}(\sigma)x^i=\frac{(1-2\sigma x(1-x))^{j-1}-1}{4\sigma j!(j-1)!}\;.
\end{equation}
The level function $H_j$ is not the reduced correlators yet but is clearly closely related. For the specified Mellin contours, we can generate the poles in $\delta$ and $\gamma$ in (\ref{defKj}) via
\begin{equation}
\nonumber \frac{1}{\delta-i}=\int_1^\infty dx x^{i-\delta-1}\;,\quad \frac{1}{\gamma-2j}=-\int_0^1 dy y^{2j-\gamma-1}\;.
\end{equation}
Introducing Schwinger parameters for the Gamma functions and performing the Mellin integrals, we get
\begin{equation}
\begin{split}
\nonumber \xi^{-1}H_j={}&-\int_1^\infty \!\!dx \int_0^1 \!\!dy\int_0^\infty da db (ab)^{j+\frac{1}{2}}y^{2j}\\
{}&\times P_j(x,\sigma)e^{-a-b-y\sqrt{ab}(\chi+\xi x)}\;.
\end{split}
\end{equation}
Note that $\xi$ and $\chi$ appear only in the exponential factor. Therefore, we define the following differential operators
\begin{equation}
\mathbb{A}=\partial_\chi^2\;,\qquad \mathbb{T}=\partial_\xi(\partial_\chi-\partial_\xi)\;,
\end{equation}
which act on the exponential as
\begin{equation}
\mathbb{A}\rightarrow aby^2\;,\quad \mathbb{T}\rightarrow aby^2 x(1-x)\;.
\end{equation}
Noting $P_2=-x(1-x)/4$, we have
\begin{equation}
\begin{split}
\frac{P_j}{P_2}{}&=\frac{1-(1-2\sigma x(1-x))^{j-1}}{\sigma x(1-x) j!(j-1)!}\\
{}&=\frac{2}{j!(j-2)!}\int_0^1 d\lambda (1-2\lambda \sigma x(1-x))^{j-2}\;.
\end{split}
\end{equation}
This allows us to rewrite $H_j$ as
\begin{equation}
H_j=\frac{2\xi}{j!(j-2)!}\int_0^1 d\lambda (\mathbb{A}-2\lambda\sigma\mathbb{T})^{j-2}f(\xi,\chi)\;,
\end{equation}
where we have defined 
\begin{equation}
f(\xi,\chi)=\frac{h_{22}}{\xi}\;.
\end{equation}
Introducing $\mathbb{B}=\mathbb{A}-2\sigma\mathbb{T}$, we then have
\begin{equation}
\mathbb{L}_\lambda=\mathbb{A}-2\lambda\sigma\mathbb{T}=(1-\lambda)\mathbb{A}+\lambda \mathbb{B}\;.
\end{equation}
Since $\mathbb{A}$, $\mathbb{B}$ commute, we can use 
\begin{equation}
\int_0^1d\lambda ((1-\lambda)\mathbb{A}+\lambda \mathbb{B})^N=\frac{1}{N+1}\sum_{r=0}^N \mathbb{A}^{N-r}\mathbb{B}^r\;,
\end{equation}
to write down an explicit differential relation for $H_j$ in terms of $H_2$
\begin{equation}
H_j=\frac{2\xi}{j!(j-1)!}\sum_{r=0}^{j-2} \mathbb{A}^{j-2-r}\mathbb{B}^r f\;.
\end{equation}

We now consider the level generator
\begin{equation}
\begin{split}
\mathcal{K}={}&\sum_{j=2}^\infty w^j H_j=2\xi w^2\int_0^1d\lambda\sum_{k=0}^\infty \frac{(w\mathbb{L}_\lambda)^k}{k!(k+2)!}f\\
={}&\xi w^2\int_0^1d\lambda\; {}_0F_1(;3; w\mathbb{L}_\lambda) f\;.
\end{split}
\end{equation}
To proceed, we make the change of variables 
\begin{equation}
U=\chi\;,\quad V=\xi+\chi\;,
\end{equation}
so that $\mathbb{L}_\lambda=(m_\lambda\partial_U+m_2\partial_V)^2$ thanks to (\ref{mlambdam2prod}).
On the other hand, we also have the identity
\begin{equation}
{}_0F_1(;3;w\mathcal E^2)=\int_{{\rm B}^2}\frac{2\,d^2p}{\pi}(1-|p|^2)e^{2\sqrt w\,p\cdot\mathcal E}\;,
\end{equation}
for any commuting two-vector of derivatives $\mathcal{E}$. Here $p$ lies in the unit disk $B^2\subset E$, and the identity follows by expanding the exponential. Identifying $\mathcal{E}$ with $m_\lambda\partial_U+m_2\partial_V$, the exponentiated action translates $U$ by $2\sqrt{w} p\cdot m_\lambda$ and $V$ by $2\sqrt{w} p\cdot m_2$. This gives
\begin{equation}
\begin{split}
\mathcal{K}={}&w^2\int_0^1d\lambda\int_{{\rm B}^2}\frac{2d^2p}{\pi}(1-|p|^2)\\
{}&\quad\quad\quad\times \frac{D_-}{\widehat{D}_-}h_{22}\left(\widehat{D}_-,\frac{\widehat{D}_+-\widehat{D}_-}{2}\right)\;,
\end{split}
\end{equation}
where we recall the relation with $D_\pm$ in (\ref{xichitoDpm}) and 
 \begin{equation}
 \widehat{D}_\pm=D_\pm+2\sqrt w\,p\cdot(m_2\pm m_\lambda)\;.
\end{equation}
The disk integral and the $\lambda$ integral combine into a weighted integral over $S^4$. In the coordinates (\ref{defp})-(\ref{deflambda}), the sphere measure is $d\Omega_4=\sqrt{1-|p|^2}\,d^2p\,d\Omega_2(u)$, with $d\Omega_2(u)=2\,d\lambda\,d\phi$. Integrating over the azimuthal angle $\phi$ and using ${\rm Vol}(S^4)=8\pi^2/3$, we obtain
\begin{equation}
\begin{split}
\nonumber{}&\int_{{\rm B}^2}\!\!\frac{2(1-|p|^2)}{\pi}d^2p\!\int_0^1 \!\!d\lambda F(p,\lambda)\\
{}&=\int_{{\rm S}^4} \frac{d\Omega_4}{{\rm Vol}({\rm S}^4)}\frac{4|\Pi_{E^\perp}\vartheta|}{3}F\left(\Pi_E\vartheta,\frac{1-\theta\cdot u(\vartheta)}{2}\right)\;,
\end{split}
\end{equation}
where we recall $|\Pi_{E^\perp}\vartheta|=\sqrt{1-|p|^2}$.
 This leads to the following key identity
\begin{equation}
\mathcal{K}=\frac{4w^2}{3}\int_{{\rm S}^4}\frac{d\Omega_4(\vartheta)}{{\rm Vol}({\rm S}^4)}|\Pi_{E^\perp}\vartheta|\frac{D_-}{\widehat{D}_-}h_{22}\left(\widehat{D}_-,\frac{\widehat{D}_+-\widehat{D}_-}{2}\right)\;.
\end{equation}

\section{Correlators of arbitrary charges}
We now need to connect the level function generator $\mathcal{K}$ to the reduced correlators $h_{k_1k_2}$. Comparing (\ref{Mellinrep}) with (\ref{defHj}), it is not difficult to see the relation 
\begin{equation}
h_{k_1k_2}=\sum_{j=2}^{k_{\rm m}}\frac{(j+1-\mathcal{T}/2)_{k_1-j}(j+1-\mathcal{T}/2)_{k_2-j}}{(k_1-j)!(k_2-j)!}H_j\;,
\end{equation}
where the differential operator
\begin{equation}
\mathcal{T}=-\xi\partial_\xi-\chi\partial_\chi\;,
\end{equation}
acts in Mellin space as multiplication by $\gamma$. Although this differential relation is obtained in Mellin space, it can be verified independently in position space. We now define 
\begin{equation}
L=(1-s)(1-t)\;,
\end{equation}
and use $\sum_{n=0}^\infty \frac{(q)_n}{n!} s^n=\frac{1}{(1-s)^q}$. For fixed $j$, the two charge sums produce $(st)^jL^{-j-1+\mathcal{T}/2} H_j$. This gives
\begin{equation}
\sum_{k_1,k_2=2}^\infty s^{k_1} t^{k_2} h_{k_1k_2}(\xi,\chi,\sigma)=\frac{1}{L}\mathcal{K}\left(\frac{st}{L};\frac{\xi}{\sqrt{L}},\frac{\chi}{\sqrt{L}},\sigma\right)\;.
\end{equation}
All in all, we have the following correlator generating function from a single $h_{22}$ correlator
\begin{equation}
\mathbf{H}=\frac{4s^2t^2}{3L^3}\!
\int_{{\rm S}^4}\!\frac{d\Omega_4(\vartheta)}{\operatorname{Vol}({\rm S}^4)}
|\Pi_{E^\perp}\vartheta|
\frac{D_-}{\widetilde D_-}
h_{22}\left(\frac{\widetilde D_-}{\sqrt L},
\frac{\widetilde D_+-\widetilde D_-}{2\sqrt L}
\right),
\end{equation}
where
\begin{equation}
\widetilde{D}_\pm=D_\pm+2\sqrt{st}\, p\cdot (m_2\pm m_\lambda)\;.
\end{equation}
All the other reduced correlators are obtained from Taylor expanding in $s$ and $t$
\begin{equation}
h_{k_1k_2}=\mathbf{H}\big|_{s^{k_1}t^{k_2}}\;.
\end{equation}

\section{Discussion and outlook}

We have presented a single-seed generating function for M-theory holographic correlators in a background that is not conformally flat. A qualitatively new feature is the weighted average over the internal sphere, replacing the familiar direct uplift of spacetime distances. It would be especially interesting to understand the origin of the shifts and averaging measure directly from eleven-dimensional supergravity.

Our result demonstrates that conformal flatness is not essential for hidden simplicity, and opens up the possibility of finding analogous generating structures in other setups. A prime target is the defect-free four-point functions of $\frac{1}{2}$-BPS operators in six-dimensional $(2,0)$ theory, which have been computed at tree level in AdS$_7\times$S$^4$ supergravity using bootstrap methods \cite{Rastelli:2017ymc,Alday:2020lbp}. Another natural setting is supergravity four-point functions in AdS$_4\times$S$^7$, for which closed-form Mellin amplitudes are known \cite{Alday:2020dtb}. It is also important to explore whether such structures persist beyond tree level. A particularly promising starting point is the surface-defect two-point functions studied here, since the one-loop correction to the lowest-weight correlator has already been computed using the defect AdS unitarity method \cite{Chen:2024orp}. It remains to be seen whether a similar generating structure can simplify the computation of higher-weight correlators at loop level. Such developments could turn the hidden simplicity uncovered here into a powerful computational tool in backgrounds where conventional hidden conformal symmetry is absent.

\vspace{0.2cm}

\noindent{\bf Acknowledgments.} We thank Junding Chen and Hynek Paul for useful comments. This work is supported by NSFC Grant No. 12275273 and funds from the Chinese Academy of Sciences.

\bibliography{refs}
\end{document}